\documentclass[english,aps,manuscript]{revtex4}
\usepackage[T1]{fontenc}
\usepackage[latin9]{inputenc}
\usepackage{amsthm}
\usepackage{amsmath}
\usepackage{amssymb}
\usepackage{graphicx}

\makeatletter
\@ifundefined{textcolor}{}
{%
 \definecolor{BLACK}{gray}{0}
 \definecolor{WHITE}{gray}{1}
 \definecolor{RED}{rgb}{1,0,0}
 \definecolor{GREEN}{rgb}{0,1,0}
 \definecolor{BLUE}{rgb}{0,0,1}
 \definecolor{CYAN}{cmyk}{1,0,0,0}
 \definecolor{MAGENTA}{cmyk}{0,1,0,0}
 \definecolor{YELLOW}{cmyk}{0,0,1,0}
 }
\numberwithin{equation}{section}
\numberwithin{figure}{section}

\makeatother

\usepackage{babel}
\begin{document}

\title{Quantum statistics of light transmitted through an intracavity Rydberg
medium}

\author{A. Grankin$^{1}$, E. Brion$^{2}$, E. Bimbard$^{1}$, R. Boddeda$^{1}$,
I. Usmani$^{1}$, A. Ourjoumtsev$^{1}$, P. Grangier$^{1}$}

\address{$^{1}$Laboratoire Charles Fabry, Institut d'Optique, CNRS, Universit\'{e}
Paris-Sud, 2 Avenue Fresnel, 91127 Palaiseau, France}

\address{$^{2}$Laboratoire Aim\'{e} Cotton, CNRS / Univ. Paris-Sud / ENS-Cachan,
B\^{a}t. 505, Campus d'Orsay, 91405 Orsay, France.}
\begin{abstract}
We theoretically investigate the quantum statistical properties of
light transmitted through an atomic medium with strong optical non-linearity
induced by Rydberg-Rydberg van der Waals interactions. In our setup,
atoms are located in a cavity and non-resonantly driven on a two-photon
transition from their ground state to a Rydberg level via an intermediate
state by the combination of the weak signal field and a strong control
beam. To characterize the transmitted light we compute the second-order
correlation function $g^{\left(2\right)}\left(\tau\right)$. The simulations
we obtained on the specific case of rubidium atoms suggest that the
bunched or antibunched nature of the outgoing beam can be chosen at
will by appropriately tuning the physical parameters. 
\end{abstract}


\maketitle

\section{Introduction}

In an optically non-linear atomic medium, dispersion and absorption
of a classical light beam depend on powers of its amplitude \cite{B08}.
At the quantum level, dispersive optical non-linearities translate
into effective interactions between photons. The ability to achieve
such strong quantum optical non-linearities is of prominent importance
in quantum communication and computation for it would allow to implement
photonic conditional two-qubit gates. The standard Kerr dispersive
non linearities obtained in non-interacting atomic ensembles, either
in off-resonant two-level or resonant three-level configurations involving
Electromagnetically Induced Transparency (EIT), are too small to allow
for quantum non-linear optical manipulations. To further enhance such
non-linearities, EIT protocols were put forward in which the upper
level of the ladder is a Rydberg level. In such schemes, the strong
van der Waals interactions between Rydberg atoms result in a cooperative
Rybderg blockade phenomenon \cite{LFC01,SWM10,CP10}, where each Rydberg
atom prevents the excitation of its neighbors inside a \textquotedbl{}blockade
sphere\textquotedbl{}. This Rydberg blockade deeply changes the EIT
profile and leads to magnified non-linear susceptibilities \cite{PMG10,DK12,PFL12,MSB13}.
In particular, giant dispersive non-linear effects were experimentally
obtained in an off-resonant Rydberg-EIT scheme using cold rubidium
atoms placed in an optical cavity \cite{PBS12,SPB13}. In this paper,
we theoretically investigate the quantum statistical properties of
the light generated in the latter protocol. Note that, contrary to
other theoretical works, e.g. \cite{GOF11,GNP13}, here, we are interested
in the dispersive regime. Moreover, since we place the atoms in a
cavity rather than in free-space, the theoretical framework and calculations
we perform also differ from \cite{GOF11,GNP13}. In particular, a
technical benefit of our approach is that we are not restricted to
considering only photon pairs but could, in principle, investigate
higher-order correlations.

We first write the dynamical equations for the system of interacting
three-level atoms coupled to the strong control field and the non-resonant
cavity mode, fed by the probe beam. We show that, under some assumptions,
the system effectively behaves as a large spin coupled to the cavity
mode \cite{GBE10}. We then compute the steady-state second-order
correlation function to characterize the emission of photons out of
the cavity. Our numerical simulations suggest that the bunched or
antibunched nature of the outgoing light as well as its coherence
time may be controlled through adjusting the detuning between the
cavity mode and probe field frequencies.

The paper is structured as follows. In Sec. \ref{sec:Theory}, we
present our setup and the assumptions we make to compute its dynamics.
We also explain the analytical and numerical methods we employ to
calculate the second-order $g^{\left(2\right)}$ correlation function
of the outgoing light beam. In Sec. \ref{sec:Numerical-results-and},
we present and interpret the results of the simulations we obtained
for $g^{\left(2\right)}\left(0\right)$ and $g^{\left(2\right)}\left(\tau>0\right)$
on the specific experimental case considered in \cite{PBS12}. Finally,
we conclude in Sec. \ref{sec:Conclusion} by evoking open questions
and perspectives of our work. Appendices address supplementary technical
details which are omitted in the text for readability.

\section{Model and methods \label{sec:Theory}}

\begin{figure}
\begin{centering}
\includegraphics[width=12cm]{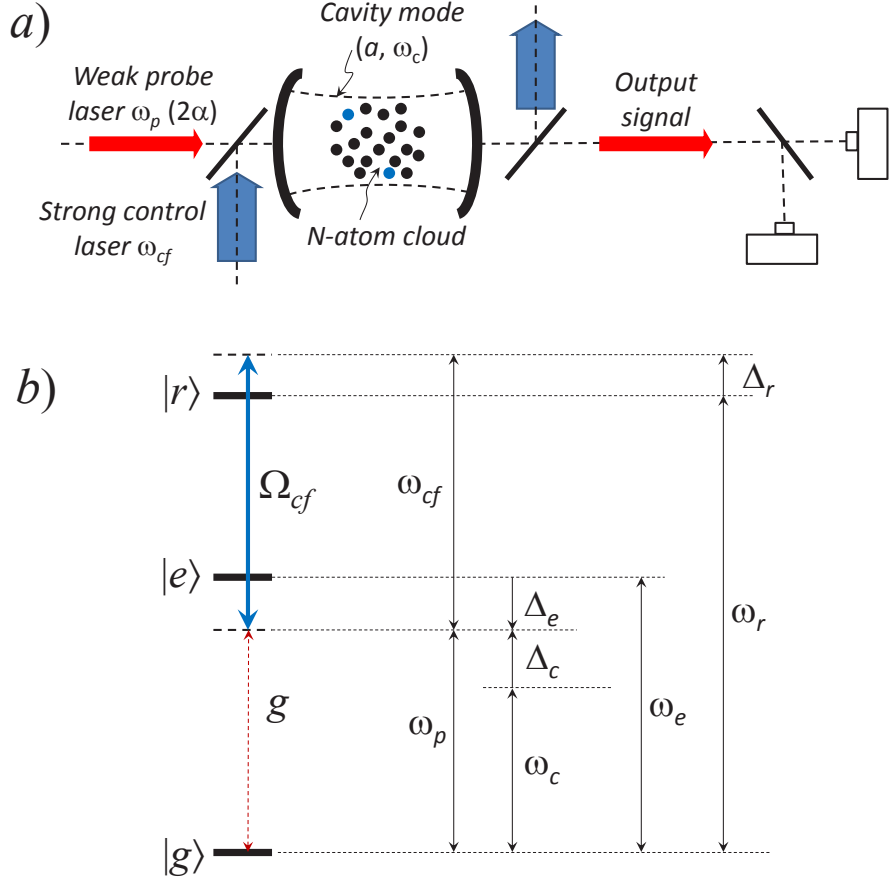} 
\par\end{centering}

\caption{a) The setup consists of $N$ cold atoms placed in an optical cavity
which is fed by a weak (classical) laser beam of frequency $\omega_{p}$
and a strong control laser field of frequency $\omega_{cf}$. b) The
atoms present a three-level ladder structure $\left\{ \left|g\right\rangle ,\left|e\right\rangle ,\left|r\right\rangle \right\} $.
The transitions $\left|g\right\rangle \leftrightarrow\left|e\right\rangle $
and $\left|e\right\rangle \leftrightarrow\left|r\right\rangle $ are
non-resonantly driven by the injected probe and control laser fields,
respectively, with the respective coupling strength and Rabi frequency
$g$ and $\Omega_{cf}$ (see the text for the definitions of the different
detunings represented here).}

\label{FigSys} 
\end{figure}

The system we consider comprises $N$ atoms which present a three-level
ladder structure with a\emph{ }ground $\left|g\right\rangle $, intermediate
$\left|e\right\rangle $ and Rydberg states $\left|r\right\rangle $
(see Fig. \ref{FigSys}). The energy of the atomic level $\left|k=g,e,r\right\rangle $
is denoted by $\hbar\omega_{k}$ (by convention $\omega_{g}=0$) and
the dipole decay rates from the intermediate and Rydberg states are
denoted by $\gamma_{e}$ and $\gamma_{r}$, respectively. The transitions
$\left|g\right\rangle \leftrightarrow\left|e\right\rangle $ and $\left|e\right\rangle \leftrightarrow\left|r\right\rangle $
are respectively driven by a weak probe field of frequency $\omega_{p}$
and a strong control field of frequency $\omega_{cf}$. To limit absorption,
both fields are off-resonant, the respective detunings are given by
$\Delta_{e}\equiv\left(\omega_{p}-\omega_{e}\right)$ and $\Delta_{r}\equiv\left(\omega_{p}+\omega_{cf}-\omega_{r}\right)$.
Moreover, to enhance dispersive effects while keeping a high input-output
coupling efficiency, the atoms are placed in an optical low-finesse
cavity. The transition $\left|g\right\rangle \leftrightarrow\left|e\right\rangle $
is supposed in the neighbourhood of a cavity resonance. The frequency
and annihilation operator of the corresponding mode are denoted by
$\omega_{c}$ and $a$, respectively ; the detuning of this mode with
the probe laser is defined by $\Delta_{c}\equiv\left(\omega_{p}-\omega_{c}\right)$
and $\alpha$ denotes the feeding rate of the cavity mode with the
probe field, which is supposed real for simplicity. Finally, we introduce
$g$ and $\Omega_{cf}$ which are the single-atom coupling constant
of the transition $\left|g\right\rangle \leftrightarrow\left|e\right\rangle $
with the cavity mode and the Rabi frequency of the control field on
the transition $\left|e\right\rangle \leftrightarrow\left|r\right\rangle $,
respectively. In the following paragraphs, we study the dynamics of
the system which, under some assumptions, is equivalent to a damped
harmonic oscillator, \emph{i.e.} the cavity mode, coupled to an assembly
of spins $\frac{1}{2}$, \emph{i.e.} the Rydberg bubbles corresponding
to the \textquotedbl{}super-atoms\textquotedbl{} delimited by the
Rydberg blockade spheres.

Starting from the full Hamiltonian, we perform the Rotating Wave Approximation
and adiabatically eliminate the intermediate state $\left|e\right\rangle $
as described in Appendix \ref{sec:Hamiltonian}. Note that the result
we obtain coincides with the lowest-order of EIT model -- the non-linearity
of the three-level atoms is neglected, and the leading non-linear
effect comes from the Rydberg-Rydberg collisional effects. The system
therefore consists of $N$ effective two-level atoms $\left\{ \left|g\right\rangle ,\left|r\right\rangle \right\} $,
with an effective power-broadened dipole decay rate from the Rydberg
level 
\[
\widetilde{\gamma}_{r}=\left(\gamma_{r}+\frac{\Omega_{cf}^{2}\gamma_{e}}{4\left(\Delta_{e}^{2}+\gamma_{e}^{2}\right)}\right),
\]
 coupled to the cavity mode of effective decay rate 
\[
\widetilde{\gamma}_{c}=\left(\gamma_{c}+\frac{g^{2}N\gamma_{e}}{\Delta_{e}^{2}+\gamma_{e}^{2}}\right)
\]
 increased by the coupling to the atomic ensemble. The Hamiltonian
reads 
\begin{eqnarray*}
\tilde{H} & = & -\hbar\tilde{\Delta}_{r}\left(\sum_{n=1}^{N}\sigma_{rr}^{\left(n\right)}\right)+\sum_{m<n=1}^{N}\hbar\kappa_{mn}\sigma_{rr}^{\left(m\right)}\sigma_{rr}^{\left(n\right)}\\
 &  & -\hbar\tilde{\Delta}_{c}a^{\dagger}a+\hbar\alpha\left(a+a^{\dagger}\right)+\hbar g_{\mathrm{eff}}\left\{ a\left(\sum_{n=1}^{N}\sigma_{rg}^{\left(n\right)}\right)+h.c.\right\} 
\end{eqnarray*}
 In this expression, we introduced the atomic operators $\sigma_{kl}^{\left(n\right)}\equiv\mathbb{I}^{\left(1\right)}\otimes\ldots\otimes\mathbb{I}^{\left(n-1\right)}\otimes\left|k\right\rangle \left\langle l\right|\otimes\mathbb{I}^{\left(n+1\right)}\otimes\ldots\otimes\mathbb{I}^{\left(N\right)}$
for $\left(k,l\right)=g,e,r$ as well as the effective detunings 
\[
\tilde{\Delta}_{r}\equiv\Delta_{r}-\frac{\Omega_{cf}^{2}\Delta_{e}}{4\left(\Delta_{e}^{2}+\gamma_{e}^{2}\right)}
\]
 and 
\[
\tilde{\Delta}_{c}\equiv\Delta_{c}-\frac{g^{2}N\Delta_{e}}{\Delta_{e}^{2}+\gamma_{e}^{2}}
\]
 respectively shifted from $\Delta_{r}$ and $\Delta_{c}$ by the
AC Stark shift of the control beam and by the linear atomic susceptibility.
The quantity $\kappa_{mn}\equiv C_{6}/\left\Vert \vec{r}_{m}-\vec{r}_{n}\right\Vert ^{6}$
is the van der Waals interaction between atoms $\left(m,n\right)$
in their Rydberg level -- when atoms are in the ground or intermediate
states, their interactions are neglected, while 
\[
g_{\mathrm{eff}}=\frac{g\Omega_{cf}}{2\Delta_{e}}
\]
 is the effective coupling strength of the two-photon transition $\left|g\right\rangle \rightarrow\left|r\right\rangle $
driven by the cavity mode and the control laser.

At this point, following \cite{GBE10}, we introduce the Rydberg bubble
approximation. In this approach, the strong Rydberg interactions are
assumed to effectively split the sample into ${\cal N}_{b}$ bubbles
$\left\{ \mathcal{B}_{\alpha=1,\ldots,\mathcal{N}_{b}}\right\} $
each of which contains $n_{b}=\left(\frac{N}{\mathcal{N}_{b}}\right)$
atoms but can only accomodate a single Rydberg excitation, delocalized
over the bubble. Note that the number of atoms per bubble $n_{b}$
is approximately given by \cite{PBS12} 
\[
n_{b}=\frac{2\pi^{2}\rho_{\mathrm{at}}}{3}\sqrt{\frac{\left|C_{6}\right|}{\Delta_{r}-\Omega_{cf}^{2}/(4\Delta_{e})}}
\]
 where $\rho_{\mathrm{at}}$ is the atomic density. Each bubble can
therefore be viewed as an effective spin $\frac{1}{2}$ whose Hilbert
space is spanned by 
\begin{eqnarray*}
\left|-_{\alpha}\right\rangle =\left|G_{\alpha}\right\rangle  & \equiv & \bigotimes_{i_{\alpha}\in\mathcal{B}_{\alpha}}\left|g_{i_{\alpha}}\right\rangle \\
\left|+_{\alpha}\right\rangle =\left|R_{\alpha}\right\rangle  & \equiv & \frac{1}{\sqrt{n_{b}}}\left\{ \left|rg\ldots g\right\rangle +\ldots+\left|g\ldots gr\right\rangle \right\} 
\end{eqnarray*}
 the ground state of the bubble $\mathcal{B}_{\alpha}$ and its symmetric
singly Rydberg excited state, respectively. Introducing the bubble
spin-$\frac{1}{2}$ operators $\mathrm{s}_{-}^{\left(\alpha\right)}=\hbar\left|-_{\alpha}\right\rangle \left\langle +_{\alpha}\right|$
-- the operator $\mathrm{s}_{-}^{\left(\alpha\right)}$ corresponds
to the lowering operator of the spin and the annihilation of a Rydberg
excitation, one can write the Hamiltonian under the approximate form
(see Appendix  \ref{sec:Hamiltonian}) 
\begin{eqnarray*}
\tilde{H} & \approx & -\hbar\tilde{\Delta}_{c}a^{\dagger}a+\hbar\alpha\left(a+a^{\dagger}\right)\\
 &  & -\hbar\tilde{\Delta}_{r}\left(\frac{\mathcal{N}_{b}}{2}+\frac{\mathrm{J}_{z}}{\hbar}\right)\\
 &  & +g_{\mathrm{eff}}\sqrt{n_{b}}\left(a\mathrm{J}_{+}+a^{\dagger}\mathrm{J}_{-}\right)
\end{eqnarray*}
 where we introduced the collective angular momentum $\mathrm{J}_{-}\equiv\sum_{\alpha=1}^{\mathcal{N}_{b}}\mathrm{s}_{-}^{\left(\alpha\right)}$.
The system is therefore equivalent to a large spin, i.e. the assembly
of spin-$\frac{1}{2}$ Rydberg bubbles, coupled to a harmonic oscillator.
Its density matrix satisfies the master equation 
\begin{eqnarray}
\partial_{t}\tilde{\rho} & = & \mathcal{L}\tilde{\rho}\label{Master}\\
 & = & \frac{1}{\mathrm{i}\hbar}\left[\tilde{H},\tilde{\rho}\right]+\widetilde{\gamma}_{c}\left\{ 2a\tilde{\rho}a^{\dagger}-a^{\dagger}a\tilde{\rho}-\tilde{\rho}a^{\dagger}a\right\} \nonumber \\
 &  & +\widetilde{\gamma}_{r}\sum_{\alpha=1}^{\mathcal{N}_{b}}\left\{ 2\mathrm{s}_{-}^{\left(\alpha\right)}\tilde{\rho}\mathrm{s}_{+}^{\left(\alpha\right)}-\mathrm{s}_{+}^{\left(\alpha\right)}\mathrm{s}_{-}^{\left(\alpha\right)}\tilde{\rho}-\tilde{\rho}\mathrm{s}_{+}^{\left(\alpha\right)}\mathrm{s}_{-}^{\left(\alpha\right)}\right\} \nonumber 
\end{eqnarray}
 One can also write the Heisenberg-Langevin equations for the time-dependent
operators $a\left(t\right),\mathrm{J}_{-}\left(t\right)$ 
\begin{eqnarray}
\partial_{t}a & = & \left(\mathrm{i}\tilde{\Delta}_{c}-\tilde{\gamma}_{c}\right)a-\mathrm{i}\alpha+\mathrm{i}g\mathrm{_{eff}}\sqrt{n_{b}}\frac{\mathrm{J}_{-}}{\hbar}+\tilde{a}_{in}\label{Heisenberg1}\\
\partial_{t}\mathrm{J}_{-} & = & \left(\mathrm{i}\tilde{\Delta}_{r}-\tilde{\gamma}_{r}\right)\mathrm{J}_{-}+\mathrm{i}\hbar g\mathrm{_{eff}}\sqrt{N\mathcal{N}_{b}}a+\tilde{\mathrm{J}}_{in}\label{Heisenberg2}
\end{eqnarray}
 where $\tilde{a}_{in},\tilde{\mathrm{J}}_{in}\equiv\sum_{n=1}^{N}\tilde{F}_{gr}^{\left(n\right)}$
are the Langevin forces associated to $a$ and $\mathrm{J}_{-}$,
respectively. Note that we neglected the effect of extra dephasing
due to, \emph{e.g.}, collisions or laser fluctuations.

To study the quantum properties of the light transmitted through the
cavity, we shall compute the function $g_{\mathrm{out}}^{\left(2\right)}$,
which characterizes the two-photon correlations. In the input-output
formalism \cite{Walls}, one shows that this function simply equals
the function $g^{\left(2\right)}$ for the intra-cavity field (see
Appendix  \ref{sec:g2} for details) given by 
\begin{equation}
g^{\left(2\right)}\left(\tau\right)=\frac{\mathrm{Tr}\left\{ a^{\dagger}ae^{\mathcal{L}\tau}\left[a\rho_{ss}a^{\dagger}\right]\right\} }{\mathrm{Tr}\left[a^{\dagger}a\rho_{ss}\right]^{2}}\label{EqG2}
\end{equation}
 where $\rho_{ss}$ denotes the steady state of the system defined
by $\mathcal{L}\rho_{ss}=0$, see Eq. (\ref{Master}).

In the regime of small feeding parameter $\alpha$, one can compute
$\rho_{ss}$ numerically by propagating in time the initial state
$\rho_{0}\equiv\left|N_{r}=0\right\rangle \left\langle N_{r}=0\right|\otimes\left|n_{c}=0\right\rangle \left\langle n_{c}=0\right|$
(here $\left|N_{r}=0,1,\ldots,\mathcal{N}_{b}\right\rangle $ represents
the symmetric state in which $N_{r}\equiv\left(\frac{\mathcal{N}_{b}}{2}+\frac{\mathrm{J}_{z}}{\hbar}\right)$
bubbles are excited, and $\left|n_{c}=0,1,\ldots\right\rangle $ are
the Fock states of the cavity mode). To this end, one applies the
Liouvillian evolution operator $e^{\mathcal{L}t}$ in a truncated
basis, restricted to states of low numbers of excitations (typically
with $n_{c}+N_{r}\leq6$). The steady state is reached in the limit
of large times -- ideally when $t\rightarrow\infty$. The denominator
of the ratio Eq.(\ref{EqG2}) is directly obtained from $\rho_{ss}$.
To compute its numerator, one first propagates in time $a\rho_{ss}a^{\dagger}$
from $t=0$ to $\tau$, using the same procedure as above, then applies
the operator $a^{\dagger}a$ and takes the trace.

In the regime of weak feeding, it is also possible to get a perturbative
expression for $g^{\left(2\right)}\left(0\right)$ by computing the
expansion of $\left\langle a^{\dagger}a^{\dagger}aa\right\rangle _{ss}$
and $\left\langle a^{\dagger}a\right\rangle _{ss}$ in powers of $\alpha$.
To this end, one uses the Heisenberg equations of the system Eqs.(\ref{Heisenberg1},\ref{Heisenberg2})
to derive the hierarchy of equations relating the different mean values
and correlations $\left\langle \ldots\right\rangle _{ss}$ up to the
fourth order in $\alpha$. After straightforward though lengthy algebra,
one gets an expression for $g^{\left(2\right)}\left(0\right)$ which
is too cumbersome to be reproduced here but allows for faster calculations
than the numerical approach. Such a fully analytical treatment, however,
cannot, to our knowledge, be extended to $g^{\left(2\right)}\left(\tau>0\right)$;
for $\tau>0$ we therefore entirely rely on numerical simulations.

To conclude this section, we consider the regime of large number of
bubbles and low number of excitations, \emph{i.e.} $\mathcal{N}_{b}\gg1$
and $\frac{\mathrm{J}_{z}}{\hbar}\ll\mathcal{N}_{b}$. As shown in
Appendix  \ref{sec:Hamiltonian}, the operator $b\equiv\frac{J_{-}}{\hbar\sqrt{\mathcal{N}_{b}}}$
is then approximately bosonic, and the term $\left(\frac{\mathcal{N}_{b}}{2}+\frac{\mathrm{J}_{z}}{\hbar}\right)$
can be put under the form 
\begin{eqnarray*}
\left(\frac{\mathcal{N}_{b}}{2}+\frac{\mathrm{J}_{z}}{\hbar}\right) & \approx & \frac{\mathrm{J}_{+}\mathrm{J}_{-}}{\hbar^{2}\left(\mathcal{N}_{b}+1\right)}+\frac{\left(\mathrm{J}_{+}\mathrm{J}_{-}\right)^{2}}{\hbar^{4}\left(\mathcal{N}_{b}+1\right)^{3}}\\
 & \approx & \frac{\mathcal{N}_{b}}{\left(\mathcal{N}_{b}+1\right)}b^{\dagger}b+\frac{\mathcal{N}_{b}^{2}}{\left(\mathcal{N}_{b}+1\right)^{3}}b^{\dagger}bb^{\dagger}b\\
 & \approx & b^{\dagger}b+\frac{1}{\mathcal{N}_{b}}b^{\dagger}b^{\dagger}bb
\end{eqnarray*}
 Finally, we get the following approximate expression for the effective
Hamiltonian 
\begin{eqnarray*}
\tilde{H} & \approx & -\hbar\tilde{\Delta}_{c}a^{\dagger}a+\hbar\alpha\left(a+a^{\dagger}\right)-\hbar\tilde{\Delta}_{r}b^{\dagger}b-\frac{\hbar\bar{\kappa}}{2}b^{\dagger}b^{\dagger}bb+\hbar g_{\mathrm{eff}}\sqrt{N}\left(ab^{\dagger}+a^{\dagger}b\right)
\end{eqnarray*}
 where $\bar{\kappa}\equiv2\tilde{\Delta}_{r}/\mathcal{N}_{b}$. In
this regime, the system therefore behaves as two coupled oscillators:
one is harmonic, the cavity field, the other is anharmonic, the Rydberg
bubble field.

In the following section, we present and discuss the results we obtained
with the specific system used in \cite{PBS12}. It appears that one
can choose the bunched or antibunched behaviour of the light transmitted
through the cavity by adjusting the detuning\textbf{ }$\Delta_{c}$.
We also show that the time behaviour of the function $g^{\left(2\right)}\left(\tau\right)$
depends on the regime considered, and can be roughly understood as
resulting from the damped exchange of a single excitation between
atoms and field.

\section{Numerical results and discussion \label{sec:Numerical-results-and}}

We consider the physical setup presented in \cite{PBS12}, \emph{i.e.}
an ensemble of $^{87}$Rb atoms, whose state space is restricted to
the levels $\left|g\right\rangle =\left|5s_{\frac{1}{2}};F=2\right\rangle $,
$\left|e\right\rangle =\left|5p_{\frac{3}{2}};F=3\right\rangle $
and $\left|r\right\rangle =\left|95d_{\frac{5}{2}};F=4\right\rangle $
with the decay rates $\gamma_{e}=2\pi\times3$ MHz, and $\gamma_{r}=2\pi\times0.03$
MHz. The other physical parameters must be designed so that strong
non-linearities may be observed at the single-photon level. In the
specific system considered here, we find this is achieved for a cavity
decay rate $\gamma_{c}=2\pi\times1$ MHz, a volume of the sample $V=40\pi\times15\times15\mu$m$^{3}$,
a sample density $n_{at}=0.4\mu$m$^{-3}$, a control laser Rabi frequency
$\Omega_{cf}=10\gamma_{e}$, a cooperativity\textbf{ $C=1000$}, a
detuning of the intermediate level $\Delta_{e}=-35\gamma_{e}$, a
detuning of the Rydberg level $\Delta_{r}=0.4\gamma_{e}$, a cavity
feeding rate $\alpha=0.01\gamma_{e}$. For these parameters, the cavity
detuning $\Delta_{c}^{\left(0\right)}=-6.1\gamma_{e}$ corresponds
to the maximal average number of photons in the cavity. Note that
these physical parameters are experimentally realistic and feasible.

\begin{figure}
\begin{centering}
\includegraphics[width=12cm]{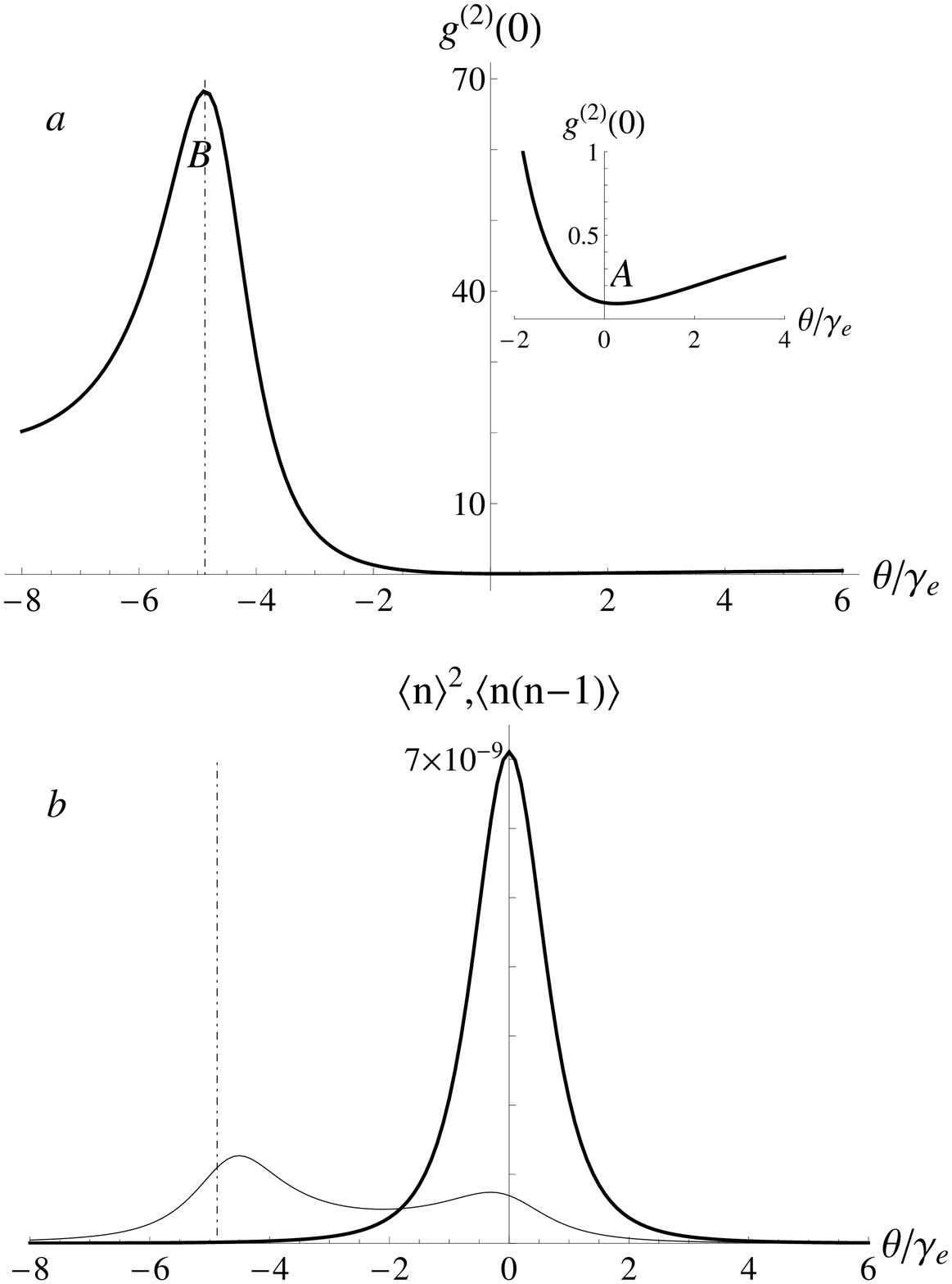} 
\par\end{centering}

\caption{a) Second-order correlation function at zero time $g^{\left(2\right)}\left(0\right)$
(numerical and analytical plots coincide), as a function of the reduced
detuning $\theta\equiv\left(\Delta_{c}-\Delta_{c}^{\left(0\right)}\right)/\gamma_{e}$.
In the neighbourhood of the minimum (A) $\theta_{A}=0$, a strong
antibunching region is observed (see inset); a strong bunching area
is obtained around the peak (B) $\theta_{B}=-4.9$. b) Average number
of pairs $\left\langle a^{\dagger}a^{\dagger}aa\right\rangle _{ss}=\left\langle n\left(n-1\right)\right\rangle _{ss}$
(thin line) and square of the average number of photons $\left\langle a^{\dagger}a\right\rangle _{ss}^{2}=\left\langle n\right\rangle _{ss}^{2}$
in the steady state (thick line). The position of the peak of the
correlation function $g^{\left(2\right)}\left(0\right)$ is signaled
by a vertical line. }

\label{FigG20} 
\end{figure}

Let us first focus on the second-order correlation function at zero
time $g^{\left(2\right)}\left(0\right)$, represented on Fig. \ref{FigG20}
a) as a function of the reduced detuning $\theta\equiv\left(\Delta_{c}-\Delta_{c}^{\left(0\right)}\right)/\gamma_{e}$.
The numerical and theoretical curves are in such a good agreement
for the regime considered that the corresponding curves cannot be
distinguished. One notes a strong bunching peak (B) $\theta_{B}=-4.9$
and a deep antibunching area centered on (A) $\theta_{A}=0$. This
suggests that around (A), photons are preferably emitted one by one,
while around (B) they are preferably emitted by pairs. Note, however,
that, as a ratio, $g^{\left(2\right)}\left(0\right)$ gives only information
on the relative importance of pair and single-photon emissions. Its
peaks therefore do not correspond to maxima of photon pair emission,
but to the best compromises between $\left\langle a^{\dagger}a^{\dagger}aa\right\rangle _{ss}$
and $\left\langle a^{\dagger}a\right\rangle _{ss}^{2}$, as can be
checked by comparison of Fig. \ref{FigG20} a) and b). Hence, pair
emission might dominate in a regime where the number of photons coming
out from the cavity is actually very small.

\begin{figure}
\begin{centering}
\includegraphics[width=12cm]{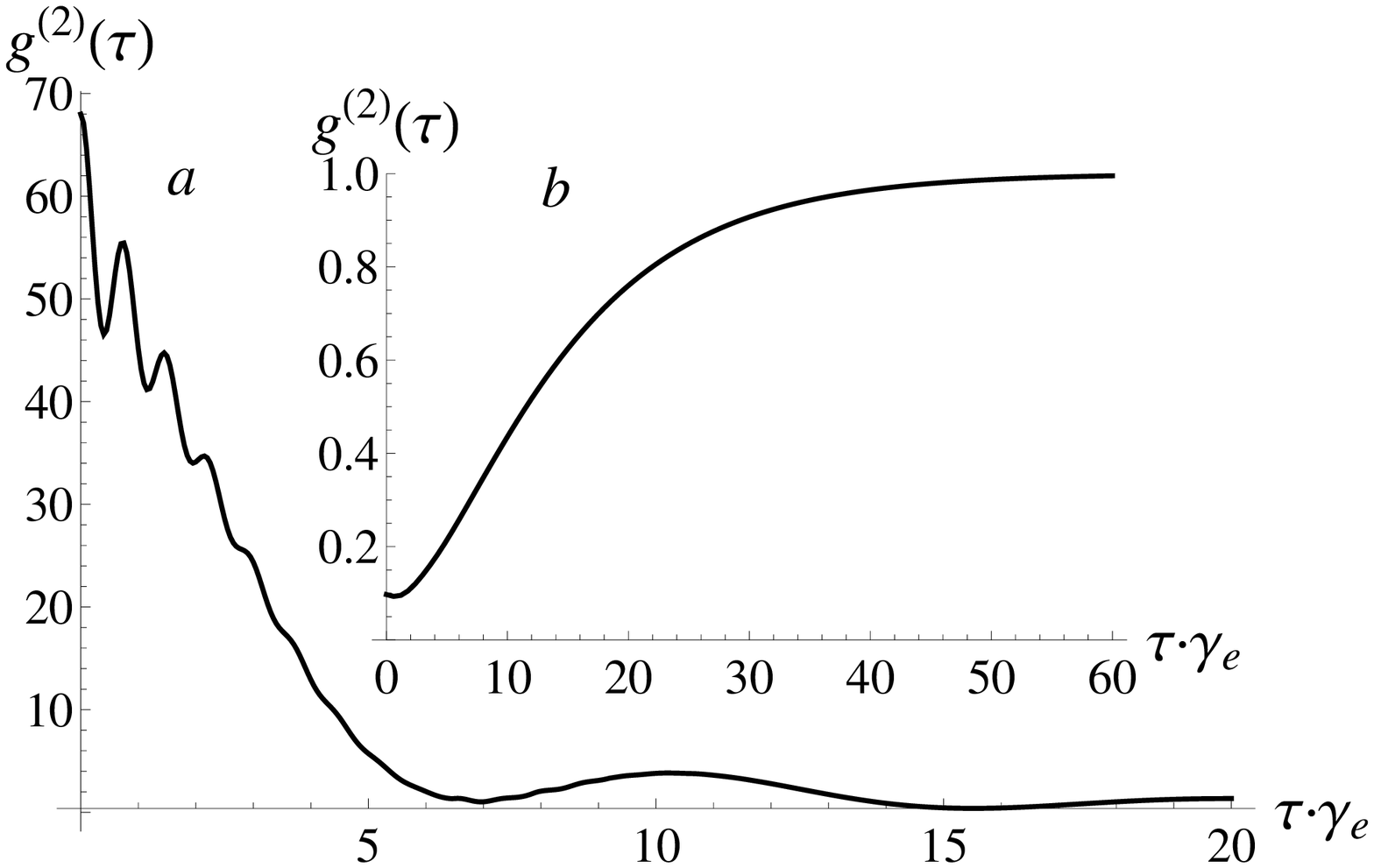} 
\par\end{centering}

\caption{Temporal behaviour of $g^{\left(2\right)}\left(\tau\right)$ for a)
$\theta_{B}=-4.9$ and b) $\theta_{A}=0$ . Note that we chose a dimensionless
{}``time''-variable $\tau\times\gamma_{e}$ on the $x$-axis. }

\label{FigG2T} 
\end{figure}

We now investigate the behaviour of $g^{\left(2\right)}\left(\tau>0\right)$
for two different values of the detuning, \emph{i.e.} $\theta_{B}=-4.9$
and $\theta_{A}=0$ which respectively correspond to the peak (B)
and minimum (A) of $g^{\left(2\right)}\left(0\right)$. The numerical
simulations we obtained are given in Fig. \ref{FigG2T}. The plot
relative to (B) exhibits damped oscillations, alternatively showing
a bunched $\left(g^{\left(2\right)}\left(\tau\right)>1\right)$ or
antibunched $\left(g^{\left(2\right)}\left(\tau\right)<1\right)$
behaviour. The plot corresponding to (A) always remains on the antibunched
side, though asymptotically tending to $1$.

The features observed can be understood and satisfactorily accounted
for by a simple three-level model. Indeed, due to the weakness of
$\alpha$, the system, in its steady state, is expected to contain
at most two excitations (either photonic or atomic). After a photon
detection at $t=0$, it contains at most one excitation which can
be exchanged between the cavity field and atoms, as it has been known
for long \cite{BOR95,BSM96}. In other words, the operator $a\rho_{ss}a^{\dagger}$
can be expanded in the space restricted to the three states $\left\{ \left|00\right\rangle \equiv\left|N_{r}=0,n_{c}=0\right\rangle ,\left|01\right\rangle \equiv\left|N_{r}=0,n_{c}=1\right\rangle ,\left|10\right\rangle \equiv\left|N_{r}=1,n_{c}=0\right\rangle \right\} $
and the effective non-Hermitian Hamiltonian for the system, in this
subspace, takes the following form: 
\[
\mathrm{H}_{3}=\hbar\left[\begin{array}{ccc}
0 & \alpha & 0\\
\alpha & -\tilde{\Delta}_{c}-\mathrm{i}\tilde{\gamma}_{c} & g_{\mathrm{eff}}\sqrt{N}\\
0 & g_{\mathrm{eff}}\sqrt{N} & -\tilde{\Delta}_{r}-\mathrm{i}\tilde{\gamma}_{r}
\end{array}\right]
\]
 The oscillatory dynamical behaviour observed for $g^{\left(2\right)}(t)$
in the specific cases (A,B) is correctly recovered by this Hamiltonian,
which validates the schematic model we used and suggests it comprises
the main physical processes at work.

To conclude this section, it is worth mentioning that the two-boson
approximation, though strictly speaking not applicable here -- the
parameters considered in this section indeed correspond to a number
of bubbles $\mathcal{N}_{b}\simeq2$, yields, however, the qualitative
behaviour for $g^{\left(2\right)}\left(0\right)$. The minimum is
correctly located, though slightly higher than in the spin model;
the antibunching peak is slightly shifted towards positive detunings
and is weaker than in the previous treatment. These discrepancies
result from too low a value of the non-linearity parameter $\bar{\kappa}$
; they can be corrected through replacing $\bar{\kappa}=2\tilde{\Delta}/\mathcal{N}_{b}$
by $\bar{\kappa}'=2\tilde{\Delta}/\left(\mathcal{N}_{b}-1\right)$
in the two-boson Hamiltonian. We first note that $\bar{\kappa}$ and
$\bar{\kappa}'$ coincide in the regime of large number of bubbles.
Moreover, $\bar{\kappa}'$ makes sense in the regime of low number
of bubbles: in particular, when $\mathcal{N}_{b}\rightarrow1$, \emph{i.e.}
when only one bubble is available, the non-linearity, proportional
to $\bar{\kappa}'$, diverges accordingly, therefore forbidding the
boson field to contain more than one excitation. Finally, let us mention
that $\bar{\kappa}'$ can also be recovered via a perturbative treatment
of the full model which will be presented in a future paper.

\section{Conclusion \label{sec:Conclusion}}

In this work, we studied how the strong Rydberg-Rydberg van der Waals
interactions in an atomic medium may affect the quantum statistical
properties of an incoming light beam. In our model, atoms are located
in a low finesse cavity and subject to a weak signal beam and a strong
control field. These two fields non-resonantly drive the transition
from the ground to a Rydberg level. The system was shown to effectively
behave as a large spin coupled to a damped harmonic oscillator, \emph{i.e.}
the assembly of Rydberg bubbles and the cavity mode, respectively.
The strong anharmonicity of the atomic spin affects the quantum statistics
of the outgoing light beam. To demonstrate this effect, we performed
analytical and numerical calculations of the second-order correlation
function $g^{\left(2\right)}\left(\tau\geq0\right)$. The results
we obtained on a specific physical example with rubidium atoms show
indeed that the transmitted light presents either bunched or antibunched
characters, depending on the detuning between the cavity mode and
the probe field. This suggests that, in such a setup, one could design
light of arbitrary quantum statistics through appropriately adjusting
the physical parameters.

In this work, we performed the Rydberg bubble approximation, which
allowed us to derive a tractable effective Hamiltonian. This scheme
is, however, questionable: interactions between bubbles are indeed
neglected, and the different spatial arrangements of the bubbles in
the sample are not considered. Though challenging, it would be interesting
to run full simulations of the system, rejecting those states which
are too far off-resonant due to Rydberg-Rydberg interactions. Besides
validating the assumption of the present work, this would indeed enable
us to consider other regimes, such as, for instance, the case of resonant
transition towards the Rydberg level. We also implicitly made the
assumption that the cavity mode and control beam were homogeneous.
Spatial variations should be included in the model and their potential
influence studied in a future work. Finally, due to the very weak
probe field regime considered in this paper, we only presented results
on the function $g^{\left(2\right)}\left(\tau\right)$: the production
of $n=3,4,\ldots$ correlated photons is indeed very unlikely. In
principle, we can, however, numerically compute $g^{\left(n\right)}\left(\tau\right)$
for any $n>2$, which might be relevant in a future work, if addressing
stronger probe fields.
\begin{acknowledgments}
This work was supported by the EU through the ERC Advanced Grant 246669
DELPHI and the Collaborative Project 600645 SIQS.
\end{acknowledgments}
\appendix

\section{Derivation of the effective Hamiltonian\label{sec:Hamiltonian}}

\subsection{Rotating Wave Approximation}

The full Hamiltonian of the system can be written under the form 
\begin{eqnarray*}
H & = & H_{a}+H_{c}+V_{a-c}\\
H_{a} & = & \hbar\omega_{e}\sum_{n=1}^{N}\sigma_{ee}^{\left(n\right)}+\hbar\omega_{r}\sum_{n=1}^{N}\sigma_{rr}^{\left(n\right)}\\
 &  & +\hbar\Omega_{cf}\cos\left(\omega_{cf}t\right)\sum_{n=1}^{N}\left(\sigma_{re}^{\left(n\right)}+\sigma_{er}^{\left(n\right)}\right)\\
 &  & +\sum_{m<n=1}^{N}\hbar\kappa_{mn}\sigma_{rr}^{\left(m\right)}\sigma_{rr}^{\left(n\right)}\\
H_{c} & = & \hbar\left[\omega_{c}a^{\dagger}a+2\alpha\cos\left(\omega_{p}t\right)\left(a+a^{\dagger}\right)\right]\\
V_{a-c} & = & \sum_{n=1}^{N}\hbar g\left(a+a^{\dagger}\right)\left(\sigma_{eg}^{\left(n\right)}+\sigma_{ge}^{\left(n\right)}\right)
\end{eqnarray*}
 where $\sigma_{\alpha\beta}^{\left(n\right)}\equiv\mathbb{I}^{\left(1\right)}\otimes\ldots\otimes\mathbb{I}^{\left(n-1\right)}\otimes\left|\alpha\right\rangle \left\langle \beta\right|\otimes\mathbb{I}^{\left(n+1\right)}\otimes\ldots\otimes\mathbb{I}^{\left(N\right)}$,
$\hbar\omega_{\alpha}$ is the energy of the atomic level $\left|\alpha\right\rangle $
for $\alpha=e,r$ (with the convention $\omega_{g}=0$), and $\kappa_{mn}\equiv\frac{C_{6}}{\left\Vert \vec{r}_{m}-\vec{r}_{n}\right\Vert ^{6}}$
denotes the van der Waals interaction between atoms in the Rydberg
level -- when atoms are in the ground or intermediate states, their
interactions are neglected.

We switch to the rotating frame defined by $\left|\psi\right\rangle \rightarrow\left|\tilde{\psi}\right\rangle =\exp\left(-\frac{\mathrm{i}t}{\hbar}H_{0}\right)$
where 
\[
H_{0}\equiv\hbar\omega_{p}a^{\dagger}a+\hbar\omega_{p}\sum_{n=1}^{N}\sigma_{ee}^{\left(n\right)}+\hbar\left(\omega_{p}+\omega_{cf}\right)\sigma_{rr}^{\left(n\right)}
\]
 and perform the Rotating Wave Approximation to get the new Hamiltonian
$\tilde{H}=\tilde{H}_{a}+\tilde{H}_{c}+\tilde{V}_{a-c}$, where 
\begin{eqnarray*}
\tilde{H}_{a} & = & -\hbar\Delta_{e}\sum_{n=1}^{N}\sigma_{ee}^{\left(n\right)}-\hbar\Delta_{r}\sum_{n=1}^{N}\sigma_{rr}^{\left(n\right)}\\
 &  & +\frac{\hbar\Omega_{cf}}{2}\sum_{n=1}^{N}\left(\sigma_{re}^{\left(n\right)}+\sigma_{er}^{\left(n\right)}\right)+\sum_{m<n=1}^{N}\hbar\kappa_{mn}\sigma_{rr}^{\left(m\right)}\sigma_{rr}^{\left(n\right)}\\
\tilde{H}_{c} & = & -\hbar\Delta_{c}a^{\dagger}a+\hbar\alpha\left(a+a^{\dagger}\right)\\
\tilde{V}_{a-c} & = & \sum_{n=1}^{N}\hbar g\left(a\sigma_{eg}^{\left(n\right)}+a^{\dagger}\sigma_{ge}^{\left(n\right)}\right)
\end{eqnarray*}
 with the detunings $\Delta_{c}\equiv\left(\omega_{p}-\omega_{c}\right)$,
$\Delta_{e}\equiv\left(\omega_{p}-\omega_{e}\right)$, and $\Delta_{r}\equiv\left(\omega_{p}+\omega_{cf}-\omega_{r}\right)$.

The corresponding Heisenberg-Langevin equations are:

\begin{eqnarray}
\frac{d}{dt}a & = & \left(\mathrm{i}\Delta_{c}-\gamma_{c}\right)a-\mathrm{i}\alpha-\mathrm{i}g\sum_{i}^{N}\sigma_{ge}^{\left(i\right)}+a_{in}\label{ApEq1}\\
\frac{d}{dt}\sigma_{ge}^{\left(i\right)} & = & \left(\mathrm{i}\Delta_{e}-\gamma_{e}\right)\sigma_{ge}^{\left(i\right)}-\mathrm{i}\frac{\Omega_{cf}}{2}\sigma_{gr}^{\left(i\right)}+\mathrm{i}ga\left(\sigma_{ee}^{\left(i\right)}-\sigma_{gg}^{\left(i\right)}\right)+F_{ge}^{\left(i\right)}\label{ApEq2}\\
\frac{d}{dt}\sigma_{gr}^{\left(i\right)} & = & \left(\mathrm{i}\Delta_{r}-\gamma_{r}\right)\sigma_{gr}^{\left(i\right)}-\mathrm{i}\frac{\Omega_{cf}}{2}\sigma_{ge}^{\left(i\right)}+\mathrm{i}ga\sigma_{er}^{\left(i\right)}\label{ApEq3}\\
 &  & -\mathrm{i}\sigma_{gr}^{\left(i\right)}\sum_{j\neq i}^{N}\kappa_{ij}\sigma_{rr}^{\left(j\right)}+F_{gr}^{\left(i\right)}\nonumber \\
\frac{d}{dt}\sigma_{er}^{\left(i\right)} & = & \left\{ \mathrm{i}\left(\Delta_{r}-\Delta_{e}\right)-\gamma_{er}\right\} \sigma_{er}^{\left(i\right)}+\mathrm{i}\frac{\Omega_{cf}}{2}\left(\sigma_{rr}^{\left(i\right)}-\sigma_{ee}^{\left(i\right)}\right)\label{ApEq4}\\
 &  & +\mathrm{i}ga^{\dagger}\sigma_{gr}^{\left(i\right)}-\,\mathrm{i}\sigma_{er}^{\left(i\right)}\sum_{j\neq i}^{N}\kappa_{ij}\sigma_{rr}^{\left(j\right)}+F_{er}^{\left(i\right)}\nonumber 
\end{eqnarray}
 where $a_{in}$ and $F_{\alpha\beta}^{\left(i\right)}$ denote Langevin
forces.

\subsection{Elimination of the intermediate state}

Let us now simplify the system. First, one deduces from Eq.(\ref{ApEq4})
that $\sigma_{er}$ is of second order in the small feeding constant
$\alpha$. The term $a\sigma_{er}^{\left(i\right)}$ can therefore
be neglected in Eq.(\ref{ApEq3}). Moreover, since the ground state
population remains dominant during the evolution of the system we
can write $\sigma_{ee}^{\left(i\right)}-\sigma_{gg}^{\left(i\right)}\simeq-\mathbb{I}$
; from Eq.(\ref{ApEq2}), the steady-state solution for $\sigma_{ge}^{\left(i\right)}$
in the far detuned regime is therefore 
\[
\sigma_{ge}^{\left(i\right)}\simeq\frac{\Omega_{cf}}{2\left(\Delta_{e}+\mathrm{i}\gamma_{e}\right)}\sigma_{gr}^{\left(i\right)}+\frac{g}{\left(\Delta_{e}+\mathrm{i}\gamma_{e}\right)}a+\frac{\mathrm{i}}{\left(\Delta_{e}+\mathrm{i}\gamma_{e}\right)}F_{ge}^{\left(i\right)}
\]
 Finally, substituting this relation into Eqs.(\ref{ApEq1},\ref{ApEq3})
one gets

\begin{eqnarray}
\frac{d}{dt}a & = & \left(\mathrm{i}\tilde{\Delta}_{c}-\tilde{\gamma}_{c}\right)a-\mathrm{i}\alpha+\mathrm{i}g\mathrm{_{eff}}\left(\sum_{i}\sigma_{gr}^{\left(i\right)}\right)+\tilde{a}_{in}\label{ApEq5}\\
\frac{d}{dt}\sigma_{gr}^{\left(i\right)} & = & \left(\mathrm{i}\tilde{\Delta}_{r}-\tilde{\gamma}_{r}\right)\sigma_{gr}^{\left(i\right)}+\mathrm{i}g\mathrm{_{eff}}a-\mathrm{i}\sigma_{gr}^{\left(i\right)}\left(\sum_{j\neq i}^{N}\kappa_{ij}\sigma_{rr}^{\left(j\right)}\right)+\tilde{F}_{gr}^{\left(i\right)}\label{ApEq6}
\end{eqnarray}
 where 
\begin{eqnarray*}
\tilde{\Delta}_{c} & = & \Delta_{c}-\Delta_{e}\frac{g^{2}N}{\left(\Delta_{e}^{2}+\gamma_{e}^{2}\right)}\\
\tilde{\gamma}_{c} & = & \gamma_{c}+\gamma_{e}\frac{g^{2}N}{\left(\Delta_{e}^{2}+\gamma_{e}^{2}\right)}\\
\tilde{\Delta}_{r} & = & \Delta_{r}-\Delta_{e}\frac{\Omega_{cf}^{2}}{4\left(\Delta_{e}^{2}+\gamma_{e}^{2}\right)}\\
\tilde{\gamma}_{r} & = & \gamma_{r}+\gamma_{e}\frac{\Omega_{cf}^{2}}{4\left(\Delta_{e}^{2}+\gamma_{e}^{2}\right)}\\
g\mathrm{_{eff}} & = & \frac{g\Omega_{cf}}{2\left(\Delta_{e}+\mathrm{i}\gamma_{e}\right)}\approx\frac{g\Omega_{cf}}{2\Delta_{e}}
\end{eqnarray*}
 are the parameters for the effective two-level model and $\tilde{a}_{in},\tilde{F}_{gr}^{\left(i\right)}$
are the modified Langevin noise operators 
\begin{eqnarray*}
\tilde{a}_{in} & = & a_{in}+\frac{g}{\left(\Delta_{e}+\mathrm{i}\gamma_{e}\right)}\sum_{i}F_{ge}^{\left(i\right)}\approx a_{in}+\frac{g}{\Delta_{e}}\sum_{i}F_{ge}^{\left(i\right)}\\
\tilde{F}_{gr}^{\left(i\right)} & = & F_{gr}^{\left(i\right)}+\frac{\Omega_{cf}}{2\left(\Delta_{e}+\mathrm{i}\gamma_{e}\right)}F_{ge}^{\left(i\right)}\approx F_{gr}^{\left(i\right)}+\frac{\Omega_{cf}}{2\Delta_{e}}F_{ge}^{\left(i\right)}
\end{eqnarray*}
 Note that, in the absence of collisional terms, one simply recovers
the standard three-level EIT susceptibility in the far-detuned regime

\[
\frac{da}{dt}=\left(\mathrm{i}\Delta_{c}-\gamma_{c}-\frac{g^{2}N}{\frac{\Omega_{cf}^{2}}{4(\gamma_{r}-\mathrm{i}\Delta_{r})}-\mathrm{i}\Delta_{e}}\right)a-\mathrm{i}\alpha+\tilde{a}_{in}
\]

Finally, we get the effective Hamiltonian 
\begin{eqnarray*}
\tilde{H} & = & -\hbar\tilde{\Delta}_{r}\left(\sum_{n=1}^{N}\sigma_{rr}^{\left(n\right)}\right)+\sum_{m<n=1}^{N}\hbar\kappa_{mn}\sigma_{rr}^{\left(m\right)}\sigma_{rr}^{\left(n\right)}\\
 &  & -\hbar\tilde{\Delta}_{c}a^{\dagger}a+\hbar\alpha\left(a+a^{\dagger}\right)+\hbar g_{\mathrm{eff}}\left\{ a\left(\sum_{n=1}^{N}\sigma_{rg}^{\left(n\right)}\right)+h.c.\right\} 
\end{eqnarray*}

\subsection{Rybderg bubble approximation}

As described in the main text, we introduce the Rydberg bubble approximation.
In this approach, the strong Rydberg interactions are assumed to effectively
split the sample into ${\cal N}_{b}$ bubbles $\left\{ \mathcal{B}_{\alpha=1,\ldots,\mathcal{N}_{b}}\right\} $
each of which contains $n_{b}=\left(\frac{N}{\mathcal{N}_{b}}\right)$
atoms but can only accomodate a single Rydberg excitation, delocalized
over the bubble. Note that the number of atoms per bubble $n_{b}$
is approximately given by \cite{PBS12} 
\[
n_{b}=\frac{2\pi^{2}\rho_{\mathrm{at}}}{3}\sqrt{\frac{\left|C_{6}\right|}{\Delta_{r}-\Omega_{cf}^{2}/(4\Delta_{e})}}
\]
 where $\rho_{\mathrm{at}}$ is the atomic density. Each bubble can
therefore be viewed as an effective spin $\frac{1}{2}$ whose Hilbert
space is spanned by 
\begin{eqnarray*}
\left|-_{\alpha}\right\rangle =\left|G_{\alpha}\right\rangle  & \equiv & \bigotimes_{i_{\alpha}\in\mathcal{B}_{\alpha}}\left|g_{i_{\alpha}}\right\rangle \\
\left|+_{\alpha}\right\rangle =\left|R_{\alpha}\right\rangle  & \equiv & \frac{1}{\sqrt{n_{b}}}\left\{ \left|rg\ldots g\right\rangle +\ldots+\left|g\ldots gr\right\rangle \right\} 
\end{eqnarray*}
 the ground state of the bubble $\mathcal{B}_{\alpha}$ and its symmetric
singly Rydberg excited state, respectively. Introducing the bubble
Pauli operators $\mathrm{s}_{-}^{\left(\alpha\right)}=\hbar\left|-_{\alpha}\right\rangle \left\langle +_{\alpha}\right|$
-- the operator $\mathrm{s}_{-}^{\left(\alpha\right)}$ corresponds
to the lowering operator of the spin and the annihilation of a Rydberg
excitation, one can write

\begin{eqnarray*}
\sum_{n=1}^{N}\sigma_{gr}^{\left(n\right)} & = & \sum_{\alpha=1}^{\mathcal{N}_{b}}\sum_{i_{\alpha}\in\mathcal{B}_{\alpha}}\sigma_{gr}^{\left(i_{\alpha}\right)}\\
 & \approx & \sum_{\alpha=1}^{\mathcal{N}_{b}}\frac{\mathrm{s}_{-}^{\left(\alpha\right)}}{\hbar}\left\langle -_{\alpha}\left|\sum_{i_{\alpha}\in\mathcal{B}_{\alpha}}\sigma_{gr}^{\left(i_{\alpha}\right)}\right|+_{\alpha}\right\rangle \\
 & \approx & \sqrt{n_{b}}\sum_{\alpha=1}^{\mathcal{N}_{b}}\frac{\mathrm{s}_{-}^{\left(\alpha\right)}}{\hbar}\\
 & = & \sqrt{n_{b}}\frac{\mathrm{J}_{-}}{\hbar}
\end{eqnarray*}
 where we introduced the collective angular momentum $\mathrm{J}_{-}\equiv\sum_{\alpha=1}^{\mathcal{N}_{b}}\mathrm{s}_{-}^{\left(\alpha\right)}$.
In the same way, 
\begin{eqnarray*}
\sum_{n=1}^{N}\sigma_{rr}^{\left(n\right)} & = & \sum_{\alpha=1}^{\mathcal{N}_{b}}\sum_{i_{\alpha}\in\mathcal{B}_{\alpha}}\sigma_{rr}^{\left(i_{\alpha}\right)}\\
 & \approx & \sum_{\alpha=1}^{\mathcal{N}_{b}}\left|+_{\alpha}\right\rangle \left\langle +_{\alpha}\right|\left\langle +_{\alpha}\left|\sum_{i_{\alpha}\in\mathcal{B}_{\alpha}}\sigma_{rr}^{\left(i_{\alpha}\right)}\right|+_{\alpha}\right\rangle \\
 & \approx & \sum_{\alpha=1}^{\mathcal{N}_{b}}\left(\frac{1}{2}+\frac{\mathrm{s}_{z}^{\left(\alpha\right)}}{\hbar}\right)\\
 & \approx & \left(\frac{\mathcal{N}_{b}}{2}+\frac{\mathrm{J}_{z}}{\hbar}\right)
\end{eqnarray*}
 where we used $\left|+_{\alpha}\right\rangle \left\langle +_{\alpha}\right|\equiv\left(\frac{1}{2}+\frac{\mathrm{s}_{z}^{\left(\alpha\right)}}{\hbar}\right)$.
Finally, the Hamiltonian of the system takes the approximate form
\begin{eqnarray*}
\tilde{H} & \approx & -\hbar\tilde{\Delta}_{c}a^{\dagger}a+\hbar\alpha\left(a+a^{\dagger}\right)\\
 &  & -\hbar\tilde{\Delta}_{r}\left(\frac{\mathcal{N}_{b}}{2}+\frac{\mathrm{J}_{z}}{\hbar}\right)\\
 &  & +g_{\mathrm{eff}}\sqrt{\mathcal{N}_{b}}\left(a\mathrm{J}_{+}+a^{\dagger}\mathrm{J}_{-}\right)
\end{eqnarray*}
 which represents the interaction of the large spin $\mathrm{J}_{-}$
with the cavity mode $a$.

\subsection{Regime of large number of bubbles and low number of excitations}

From the well-known relation $\mathrm{J}_{+}\mathrm{J}_{-}=\vec{\mathrm{J}}^{2}-\mathrm{J}_{z}^{2}+\hbar\mathrm{J}_{z}$
we deduce the second-order operator equation 
\[
\mathrm{J}_{z}^{2}-\hbar\mathrm{J}_{z}-\hbar^{2}\frac{\mathcal{N}_{b}}{2}\left(\frac{\mathcal{N}_{b}+2}{2}\right)+\mathrm{J}_{+}\mathrm{J}_{-}=0
\]
 In the regime of large number of bubbles $\mathcal{N}_{b}\gg1$ and
for low excitation numbers, \emph{i.e.} eigenstates of the total angular
momentum $\left|j=\frac{\mathcal{N}_{b}}{2};m=-\frac{\mathcal{N}_{b}}{2}+k\right\rangle $
with $k\ll\mathcal{N}_{b}$, the solution of this equation is approximately
given by 
\[
\mathrm{J}_{z}\approx\hbar\left\{ -\frac{\mathcal{N}_{b}}{2}+\frac{\mathrm{J}_{+}\mathrm{J}_{-}}{\hbar^{2}\left(\mathcal{N}_{b}+1\right)}+\frac{\left(\mathrm{J}_{+}\mathrm{J}_{-}\right)^{2}}{\hbar^{4}\left(\mathcal{N}_{b}+1\right)^{3}}\right\} 
\]
 whence, at the lowest order in the excitation number, 
\begin{eqnarray}
\left(\frac{\mathcal{N}_{b}}{2}+\frac{\mathrm{J}_{z}}{\hbar}\right) & \approx & \frac{\mathrm{J}_{+}\mathrm{J}_{-}}{\hbar^{2}\left(\mathcal{N}_{b}+1\right)}+\frac{\left(\mathrm{J}_{+}\mathrm{J}_{-}\right)^{2}}{\hbar^{4}\left(\mathcal{N}_{b}+1\right)^{3}}\label{nonlinearity-1}\\
\left[\mathrm{J}_{+},\mathrm{J}_{-}\right] & \approx & -\hbar^{2}\mathcal{N}_{b}\label{bosonic-1}
\end{eqnarray}

Injecting Eq.(\ref{nonlinearity-1}) into the previous form of the
Hamiltonian we get 
\begin{eqnarray*}
\tilde{H} & \approx & -\hbar\tilde{\Delta}_{c}a^{\dagger}a+\hbar\alpha\left(a+a^{\dagger}\right)\\
 &  & -\hbar\tilde{\Delta}_{r}\left(\frac{\mathrm{J}_{+}\mathrm{J}_{-}}{\hbar^{2}\left(\mathcal{N}_{b}+1\right)}+\frac{\left(\mathrm{J}_{+}\mathrm{J}_{-}\right)^{2}}{\hbar^{4}\left(\mathcal{N}_{b}+1\right)^{3}}\right)\\
 &  & +\hbar g_{\mathrm{eff}}\sqrt{N}\left(a\frac{\mathrm{J}_{+}}{\hbar\sqrt{\mathcal{N}_{b}}}+a^{\dagger}\frac{\mathrm{J}_{-}}{\hbar\sqrt{\mathcal{N}_{b}}}\right)
\end{eqnarray*}
 Moreover, from Eq.(\ref{bosonic-1}) we deduce that the operator
$b\equiv\frac{J_{-}}{\hbar\sqrt{\mathcal{N}_{b}}}$ is approximately
bosonic and therefore, the Hamiltonian can finally be put under the
form 
\begin{eqnarray*}
\tilde{H} & \approx & -\hbar\tilde{\Delta}_{c}a^{\dagger}a+\hbar\alpha\left(a+a^{\dagger}\right)-\hbar\tilde{\Delta}_{r}b^{\dagger}b-\frac{\hbar\bar{\kappa}}{2}b^{\dagger}b^{\dagger}bb+\hbar g_{\mathrm{eff}}\sqrt{N}\left(ab^{\dagger}+a^{\dagger}b\right)
\end{eqnarray*}
 where $\bar{\kappa}\equiv2\tilde{\Delta}_{r}/\mathcal{N}_{b}$.

\section{Calculation of $g_{\mathrm{out}}^{\left(2\right)}$\label{sec:g2}}

By definition, the second-order correlation function for the outgoing
field is

\[
g_{out}^{\left(2\right)}(t_{1},t_{2})=\frac{\left\langle a_{out}^{\dagger}\left(t_{1}\right)a_{out}^{\dagger}\left(t_{2}\right)a_{out}\left(t_{2}\right)a_{out}\left(t_{1}\right)\right\rangle }{\left\langle a_{out}^{\dagger}\left(t_{2}\right)a_{out}\left(t_{2}\right)\right\rangle \left\langle a_{out}^{\dagger}\left(t_{1}\right)a_{out}\left(t_{1}\right)\right\rangle }
\]
 Using the relations \cite{Walls} 
\begin{eqnarray*}
\left\langle a_{out}^{\dagger}\left(t\right)a_{out}\left(t\right)\right\rangle  & = & 2\gamma_{c}\left\langle a^{\dagger}\left(t\right)a\left(t\right)\right\rangle \\
a_{out}\left(t\right) & = & \sqrt{2\gamma_{c}}a\left(t\right)-a_{in}\left(t\right)
\end{eqnarray*}
 and keeping only non-zero terms (all terms like $\left\langle a_{in}^{\dagger}....\right\rangle $
and $\left\langle ....a_{in}\right\rangle $ equal zero), one obtains
in the numerator four non-zero terms

\begin{eqnarray*}
\left\langle a^{\dagger}(t_{1})a^{\dagger}(t_{2})a_{in}(t_{2})a(t_{1})\right\rangle \\
\left\langle a^{\dagger}(t_{1})a^{\dagger}(t_{2})a(t_{2})a(t_{1})\right\rangle \\
\left\langle a^{\dagger}(t_{1})a_{in}^{\dagger}(t_{2})a(t_{2})a(t_{1})\right\rangle \\
\left\langle a^{\dagger}(t_{1})a_{in}^{\dagger}(t_{2})a_{in}(t_{2})a(t_{1})\right\rangle 
\end{eqnarray*}

Let us consider the first term. Using the standard commutation relations
between $a$ and $a_{in}$ operators we have:

\begin{eqnarray*}
\left\langle a^{\dagger}(t_{1})a^{\dagger}(t_{2})a_{in}(t_{2})a(t_{1})\right\rangle  & = & \left\langle a^{\dagger}(t_{1})a^{\dagger}(t_{2})a(t_{1})a_{in}(t_{2})\right\rangle \\
 &  & +\left\langle a^{\dagger}(t_{1})a^{\dagger}(t_{2})\left[a_{in}(t_{2}),a(t_{1})\right]\right\rangle \\
 & = & \sqrt{2\gamma_{c}}\theta(t_{1}-t_{2})\left\langle a^{\dagger}(t_{1})a^{\dagger}(t_{2})\left[a(t_{2}),a(t_{1})\right]\right\rangle 
\end{eqnarray*}

Here we used the relation 
\[
\left[X\left(t_{1}\right),a_{in}\left(t_{2}\right)\right]=\sqrt{2\gamma_{c}}\theta\left(t_{1}-t_{2}\right)\left[X,a\right]
\]
 where $X$ is any system operator \cite{Walls} and where $\theta\left(\tau\right)$
is the Heaviside step-function (with $\theta\left(0\right)=\frac{1}{2}$).

Evaluating the other terms in the same way one finally obtains

\[
g_{out}^{(2)}\left(t_{1},t_{2}\right)=\frac{\left\langle a^{\dagger}\left(t_{m}\right)a^{\dagger}\left(t_{M}\right)a\left(t_{M}\right)a\left(t_{m}\right)\right\rangle }{\left\langle a^{\dagger}\left(t_{1}\right)a\left(t_{1}\right)\right\rangle \left\langle a^{\dagger}\left(t_{2}\right)a\left(t_{2}\right)\right\rangle }
\]
 where $t_{m}\equiv\mathrm{min}\left(t_{1},t_{2}\right)$ and $t_{M}\equiv\mathrm{max}\left(t_{1},t_{2}\right)$.

\end{document}